\documentclass{amsart}
\usepackage{graphicx}
\usepackage{epsf}

\usepackage{amsmath, amssymb,latexsym,float,algorithm,algorithmic}
\usepackage{enumerate}

\theoremstyle{definition}

\theoremstyle{remark}

\numberwithin{equation}{section}
\newcommand{\norm}[1]{\left\Vert#1\right\Vert}
\newcommand{\abs}[1]{\left\vert#1\right\vert}

\newcommand{\Real}{\mathbb R}
\newcommand{\Complex}{\mathbb C}

\def\la{\langle}
\def\ra{\rangle}

\newcommand{\R}{\text{\fontshape{n}\selectfont I\kern-.42exR}}

\newcommand{\1}{\text{\fontshape{n}\selectfont 1\kern-.56exl}}

\begin{document}

\centerline{\Large\it UKQCD Collaboration and University of Edinburgh}

\title[Lattice QCD with Suppressed High Momentum Modes]
{Lattice QCD with Suppressed High Momentum Modes
of the Dirac Operator}

\author{Artan Bori\c{c}i}



\date{26 Aug 2002}
\maketitle
\begin{abstract}
I define lattice fermions in five Euclidean dimensions and
the corresponding effective theory in four dimensions.
The main properties of these theories include the suppression of high
momentum modes of the lattice Dirac operator and their
ability to continuously interpolate between quenched and
dynamical fermions. In particular, the standard formulation of
lattice QCD can be viewed as a limiting case of the theory.
\end{abstract}
\pagebreak

Lattice simulations are an indispensable tool in understanding
the strong force. Since its formulation by
\cite{Wilson}, lattice Quantum Chromodynamics (QCD) has progressed
into a separate discipline.

Nevertheless, simulations of lattice QCD are still away form the
precision tests that one would like to see.
Clearly, faster computers and simulation algorithms are needed.

Recently, the issue of `ultraviolet (UV) slowing down' has attracted
particular interest
\cite{Irving_et_al,Duncan_et_al,PhdF,Peardon,AHasenfratz_Knechtli}.
These studies try to address algortmically large fluctuations
of the high end modes of the fermion determinant.
The goal of these algorithms is to
separate the UV modes and to focus on the silumation of the
physically interesting infrared modes.
In this paper I show that all this computational effort can be saved
if one improves UV properties of the lattice Dirac operator
in the first place.
 
Much of UV slowing down is thought to come from the
non-smoothness of the gauge fields \cite{AHasenfratz_Knechtli}.
The effects of non-smooth gauge fields are mostly observed in the
large eigenvalues of the fermion determinant \cite{Irving_et_al}.
Recently, \cite{Duncan_et_al} proposed a strategy to accelerate
fermion simulations by computing directly the `infrared'
eigenvalues and attaching the `ultraviolet' ones by
approximate actions and the multiboson method. \cite{PhdF}
proposed an algorithm which `filters' these fermion modes.
Another strategy is the inclusion of
ultraviolet modes in a multibosonic fashion which results
in faster algorithms \cite{Peardon}.
A direct smoothing approach is also possible
by smearing techniques \cite{AHasenfratz_Knechtli}.

In spite of the recent progress there is no unifying view
how to deal with the UV slowing down.
In this paper I propose an improved formulation of lattice
fermions which
`quenches' fermion eigenvalues beyond a given eigenvalue level.
The paper proves rigorously the existence of such a field theory.

In the following section I discuss the need to deal with
the problem of non-smooth lattice gauge fields.
In section 2 I give a model of five dimensional
fermions with suppressed high fermion modes.
I then define the theory in four dimensions and
prove its the field theoretic properties in section 3.
Finally, in section 4 I draw the conclusions.

\section{Difficulties with lattice fermions}

\subsection{Basic definitions}

The lattice regularization of gauge theories was defined
by \cite{Wilson}.

A fermion field on a regular Euclidean lattice
$\Lambda$ is a Grassmann valued vector
$\psi(x), ~~~x=\{ x_{\mu},\mu=1,\ldots,4 \} \in \Lambda$
which carries spin and color indices.
The first and second order differences
are defined by the following expressions:

\begin{equation*}
\begin{array}{l}
{\hat \partial}_{\mu} \psi(x) = \frac{1}{2 a} [\psi(x + a e_{\mu}) -
\psi(x - a e_{\mu})] \\
\\
{\hat \partial}_{\mu}^2 \psi(x) = \frac{1}{~~~~a^2} [\psi(x + a e_{\mu}) +
\psi(x - a e_{\mu}) - 2 \psi(x)]
\end{array}
\end{equation*}
where $a$ and $e_{\mu}$ are the lattice spacing and the unit
lattice vector along the coordinate $\mu=1,\ldots,4$.
Let $U(x)_{\mu} \in \Complex^{3\times 3}$ be
an element of the $SU(3)$ group, the oriented link
connecting lattice sites $x$ and $x + a e_{\mu}$. Then covariant
differences are defined by:

\begin{equation*}
\begin{array}{l}
\nabla_{\mu} \psi(x) = \frac{1}{2 a} [U(x)_{\mu} \psi(x + a e_{\mu}) -
U^{\dag}(x - a e_{\mu})_{\mu} \psi(x - a e_{\mu})] \\
\\
\Delta_{\mu} \psi(x) = \frac{1}{~~~~a^2} [U(x)_{\mu} \psi(x + a e_{\mu}) +
U^{\dag}(x - a e_{\mu})_{\mu} \psi(x - a e_{\mu}) - 2 \psi(x)]
\end{array}
\end{equation*}
The Wilson-Dirac operator is a matrix operator
$D_W \in \Complex^{N\times N}$ defined by:
\begin{equation}\label{Wilson_op}
D_W = m \1 + \sum_{\mu = 1}^4 (\gamma_{\mu} \nabla_{\mu} -
\frac{a}{2} \Delta_{\mu})
\end{equation}
where $\1$ is the identity matrix and $m$ the bare mass of the
fermion;
$\{ \gamma_{\mu} \in \Complex^{4\times 4},\mu=1,\ldots,4 \}$
is the set of
anti-commuting and Hermitian gamma-matrices of the Dirac-Clifford
algebra.
The fermion lattice action is defined by:
\begin{equation}
S_f = \sum_{x,y \in \Lambda} {\bar \psi}(x) D_W(x,y) \psi(y)
\end{equation}
whereas the gauge action is given by:
\begin{equation}
S_g = \frac{1}{g^2} \sum_{\mathbb{P}} (\1 - U_{\mathbb{P}})
\end{equation}
The sum in the right hand side
is over all plaquettes $\mathbb{P}$ on the lattice.
$U_{\mathbb{P}}$ is $1\times 1$ Wilson loop and $g$ is the bare
coupling constant of the theory.

The basic computational task in lattice QCD is the evaluation of
the partition function given by:
\begin{equation}\label{Z_QCD}
Z_{QCD} = \int \sigma_H(U) \sigma(\psi,{\bar \psi}) e^{-S_f - S_g}
\end{equation}
where $\sigma_H(U)$ and $\sigma(\psi,{\bar \psi})$ denote
the Haar and Grassmann measures respectively.
The computing problem has $O(e^N)$ complexity (i.e. it is NP-hard)
and one has to resort to stochastic estimations of the right hand side
(\ref{Z_QCD}).
In fact,
integration over the Grassmann fields can be performed exactly to
give:
\begin{equation*}
Z_{QCD} = \int \sigma_H(U) \det D_W e^{-S_g}
\end{equation*}

\subsection{A measure of non-smoothness of lattice gauge fields}

I will use a linear function of plaquette
to characterize the degree of the non-smoothness on the lattice.
Let $P(U)$ be the average $1\times 1$ Wilson loop over all lattice plaquettes.
Then I define the following real and positive function:
\begin{equation}\label{non_smoothness}
\Phi(U) = 1 - P(U) = 1 - \la U_{\mathbb{P}} \ra_{\mathbb{P}}
\end{equation}
Smooth gauge fields are characterized by
small values of $\Phi(U)$. Typical values from lattice simulations
are $\Phi(U) \in (0.4,0.5)$, which is a clear indication of
non-smoothness of the lattice gauge fields.

To see the influence of the non-smoothness on a typical gauge invariant
lattice operator, I consider eigenvalue perturbations of the
lattice Dirac operator.
Let $\lambda(U)$ be an eigenvalue of the matrix $A(U) \equiv D_W^{\dag}D_W$.
A classical result from the eigenvalue perturbation theory states that
variation of an eigenvalue $\delta\lambda(U)$ under matrix perturbation
$\delta A(U)$ is bounded by \cite{Golub_VanLoan}:
\begin{equation}\label{eig_pert}
\abs{\delta\lambda(U)} \leq \norm{\delta A(U)}_p, ~~~~~p \in \mathbb{N}
\end{equation}

The following result relates the 2-norm of the matrix $A(U)$ to the
degree of the non-smoothness $\Phi(U)$ on the lattice (\ref{non_smoothness}):
\begin{equation}\label{Appendix_A}
\norm{\delta A(U)}_2 \leq c_1 + c_2 \sqrt{\Phi(U)}
\end{equation}
where $c_1$ and $c_2$ are positive constants.
This result is proven in Appendix A. Together with the bound (\ref{eig_pert})
it gives:
\begin{equation}\label{eig_pert2}
\abs{\delta\lambda(U)} \leq c_1 + c_2 \sqrt{\Phi(U)}
\end{equation}              

The above inequality (\ref{eig_pert2}) states that the
eigenvalue fluctuations on a lattice with smooth gauge fields
are likely to be smaller than those on a lattice
with non-smooth gauge fields.

However, as present computing resources do not allow a close
approach to continuum limit, {\it it is important
to look for formulations and algorithms with reduced
effects of lattice non-smoothness.}

Numerical results of \cite{Duncan_et_al} indicate
that large eigenvalues of the lattice Dirac operator are merely
lattice artifacts.
It is a well-known fact that cutoff modes
are poorly represented on the lattice.
But one may not simply exclude them
from simulations since the existence
of the theory may be compromised.
The strategy followed in this paper
is the suppression of the high fermion modes of the lattice theory.
As shown below it is possible to model a
fermion theory with the reduced appearance of these modes.

\section{Modeling cutoff modes: Wilson fermions in 4$+$1 Euclidean dimensions}

Recent progress with chiral fermions on the lattice has shown that
a theory of five dimensional fermions can be a useful modeling
tool (for a review see \cite{Kikukawa}). 
Domain wall boundary conditions along the fifth dimension
provide a kinematical model for QCD with chiral fermions which are
``localized'' on the surface of a five dimensional world.
The theory in five dimensions can be viewed as a fermion system
propagating along the fifth Euclidean dimension with its dynamics
generated by a certain Hamiltonian operator $\mathcal{H}$.

Let $c^{\dag}(x)$ and $c(x)$ be creation and destruction operators
in the Fock space which satisfy the anti-commutation relations:
\begin{equation*}
[c^{\dag}(x),c(x)]_+ = \1
\end{equation*}
They carry spin and color indices, which are not explicitly shown
for clarity. $c(x)$ acts on the bare vacuum state which they
annihilate.

I define the Hamiltonian operator of the fermion system by the
bilinear form:
\begin{equation*}
\mathcal{H} = \sum_{x,y \in \Lambda} c^{\dag}(x) H_W(x,y) c(y)
\end{equation*}
where $H_W$ is the Hermitian lattice Dirac operator in four
dimensions given by:
\begin{equation*}
H_{W} = \gamma_5 D_{W}
\end{equation*}

Let $L_5$ be the lattice size in the fifth dimension or the ``the
inverse temperature'' of the quantum statistical system with the
partition function $Z$, given by:
\begin{equation*}
Z = \text{Tr}~~ e^{-L_5 \mathcal {H}}
\end{equation*}

To compute the trace of an operator in the Fock space I use the
standard technique of the Grassmann coherent states with
antiperiodic boundary conditions in the fifth Euclidean
coordinate. Since $\mathcal{H}$ is quadratic it is easy to show
that:
\begin{equation*}
Z = \det(\1 + e^{-L_5 H_W})
\end{equation*}

I {\it define} the measure density of the non-trivial fermion
theory on the lattice by the following equation:
\begin{equation*}
\omega(H_W) = \frac{1}{Z} \text{Tr}~~ \mathcal{O} e^{-L_5\mathcal{H}}
\end{equation*}
where $\mathcal{O}$ is a function of the creation and annihilation
operators. I choose it such that the right hand side is given by:
\begin{equation*}
\omega(H_W) = \frac{1}{Z} \det(\1 - e^{-L_5 H_W})
\end{equation*}
Such an operator exists as it is shown in the
framework of domain wall fermions \cite{Furman_Shamir}.

Hence the resulting density can be written as:
\begin{equation}\label{omega}
\omega(H_W) = \det(\tanh \frac{L_5}{2} H_W)
\end{equation}

This suggests that an effective theory in four dimensions may be
defined by the following lattice Dirac operator:
\begin{equation}\label{D_operator}
D = \frac{\mu}{a} \gamma_5 \tanh \frac{a H_W}{\mu}
\end{equation}
where $a$ is the lattice spacing of the four dimensional lattice
and $\mu > 0$ is a dimensionless parameter.
It is clear that for small lattice spacing this operator
approaches the Wilson Dirac operator and hence has the correct
continuum limit.

In order to give to the Tr$~~\mathcal{O}$ operator a precise meaning, I {\it
define} a fermion theory in five dimensions by the following
action:
\begin{equation}\label{S_5}
S^{(5)} = {\bar \psi}^{(5)} D_W^{(5)} \psi^{(5)}
\end{equation}
Here the five dimensional fermion field $\psi^{(5)}$ satisfies
periodic boundary conditions in all directions and $D_W^{(5)}$ is defined by:
\begin{equation*}
D_W^{(5)} = D_W + \gamma_5 \nabla_5 - \frac{a_5}{2} \Delta_5
\end{equation*}

It can be shown that:
{\it The measure density of the
five dimensional theory (\ref{S_5})
is proportional to the measure of the
effective theory defined by eq. (\ref{omega})}.
The proof is given in Appendix B.

This result suggests that the four
dimensional lattice theory with Wilson fermions can be approached
by the `high temperature' limit of a theory with Wilson fermions
in five dimensions.
Thus, it is natural to choose the length of the extra
dimension to be proportional to the lattice spacing in four
dimensions. Dimensional reduction is then realized by taking
the continuum limit of the theory.
Furthermore, the theory allows
the introduction of a dimensionless parameter $\mu$
which can be used to suppress the high momentum modes of the
fermion theory to a prescribed level, i.e. $\mu$ can be viewed as
a dimensionless `temperature'. A `cold' theory would then
correspond to the quenched approximation,
whereas a `hot' one would be identical to the Wilson theory.

In the next section I will give the basic properties of the
dimensionally reduced effective theory.

\section{A fermion theory with suppressed cutoff modes}

It is clear that the above theory defined in 4$+$1 dimensions
has all desired properties of a field theory: it is local,
unitary and gauge invariant.

This is not obvious for the  dimensionally reduced theory
with a lattice Dirac operator given by (\ref{D_operator}):
\begin{equation}
D = \frac{\mu}{a} \gamma_5 \tanh \frac{a H_W}{\mu}
\end{equation}
For free fermions the lattice Dirac
operator has a momentum space representation given by:
\begin{equation}\label{inverse_propagator}
{\tilde D}(p) = \frac{\mu}{a} \gamma_5 \tanh \frac{a {\tilde
H_W}(p)}{\mu}
\end{equation}
with $p = \{p_{\nu}, \nu=1,\ldots,4\}$ being the four-momentum
vector.
The momentum space Wilson Dirac operator is given by:
\begin{equation*}
{\tilde D}_W(p) = \frac{1}{a} \sum_{\nu=1}^4(\1 - e^{-i
\gamma_{\nu} ap_{\nu}})
\end{equation*}
whereas its square is given by:
\begin{equation*}
a^2 {\tilde D}_W(p)^{\dag} {\tilde D}_W(p) = \sum_{\nu=1}^4 \sin^2
ap_{\nu} + [\sum_{\nu=1}^4(1 - \cos^2 ap_{\nu})]^2
\end{equation*}

\subsection{Locality}

Whilst $D_W$ links only nearest neighbour lattice points, $D$
will be a full matrix. A full matrix can be considered
essentially local if it is dominated by matrix elements
which link lattice points that are close to each other.
For example, this will be the case
if the magnitude of $D_{ij}$
decays exponentially with the distance $|i - j|$.
This will be considered as a
sufficient condition in the following for locality
(see also \cite{Hernandes_et_al}).
Since ${\tilde D}(p)$ is analytic and $2 \pi$-periodic for $\mu >
0$, then its Fourier transform falls off exponentially
at large distances (see \cite{Luscher98} for a similar argument).
Therefore, $D$ {\it is a local operator in the above sense}.

The locality of the fermion theory in a gauge field background
is treated in Appendix C. In particular, it is shown that
if the Wilson Dirac operator is singular the locality of the theory
is guaranteed solely by the positivity of $\mu$.

\subsection{Unitarity}

For unitarity it is sufficient to show that
the lattice operator leads to non-negative energy spectrum with
non-negative norm of eigenmodes.
To do this I define a positive function $\tilde f$ in terms of the
real variable $z$:
\begin{equation*}
{\tilde f}(z) = \frac{1}{z} \tanh z, ~~~~z \neq 0
\end{equation*}
Since $\tanh z$ is an odd function of $z$, one can easily show
that the right hand side is in fact a function of $z^2$ only.
Therefore, I can define a function $f$ such that:
\begin{equation}\label{f_definition}
\begin{array}{l}
f(z^2) = \frac{1}{z} \tanh z, ~~~~z \neq 0 \\
f(0) = 1
\end{array}
\end{equation}
This way, I may write:
\begin{equation*}
\tanh z = z f(z^2), ~~~~z \in \R
\end{equation*}

Using the definition of an operator-valued function
the lattice Dirac operator (\ref{D_operator}) takes the
form:
\begin{equation}\label{D_form_1}
D = D_W f[(\frac{a H_W}{\mu})^2]
\end{equation}

Now note that the matrix function $f(.)$ is non-singular. Hence the
poles in the fermion propagator are identical to those in the
Wilson theory and the resulting theory is characterized by a real
energy spectrum. Moreover, since $f(.)$ is positive definite
the norm of energy eigenmodes remains positive.

\subsection{Perturbation theory}

The fermion propagator is given by the inverse of the
expression (\ref{inverse_propagator}).
As usual, gauge fields are parametrized by $su(3)$
elements:
\begin{equation}\label{su3}
U(x)_{\mu} = e^{i a g A(x)_{\mu}}, ~~~~A(x)_{\mu} \in su(3)
\end{equation}
and the Wilson operator is written as a sum of
the free and interaction terms:
\begin{equation*}
D_W = D_W^0 + D_W^I
\end{equation*}
The splitting of the lattice Dirac operator is written
in the same form:
\begin{equation*}
D = D^0 + D^I,
~~~~D^0 = \frac{\mu}{a} \gamma_5 \tanh \frac{a H_W^0}{\mu}
\end{equation*}
where the interaction term has to be determined.
This can be done by expanding $D$ in terms of $a/\mu$:
\begin{equation}\label{pert_expan}
D = D_W[\1 + c_1 (\frac{a H_W}{\mu})^2
           + c_2 (\frac{a H_W}{\mu})^4 + \cdots]
\end{equation}
where $c_1, c_2, \ldots$ are real expansion coefficients.

Calculation of $D^I$ is outside of the scope of this paper.
In fact it is an easy task if one stays with a finite
number of terms in the right hand side of (\ref{pert_expan}).
Also, the number of terms can be minimized using a
Chebyshev approximation for the hyperbolic tangent.
\footnote{I would like to thank Joachim Hein for discussions
related to lattice perturbation theory.}

\subsection{Fixing $\mu$}

Tuning $\mu$ and then fixing it to a certain value
is essential in specifying the level of UV suppression and the
fermion theory as such.
On the practical side one should know how to choose the value
of $\mu$ such that only the infrared modes are included.
This can be done by computing explicitly the eigenvalue density of the
input operator $H_W$ and identify the threshold between the
physical modes and the tail of the distribution.
Following definition (\ref{Wilson_op}) and the technique of \cite{EHN98}
one can approximate the density of zero eigenvalues
$\rho(0,m)$ of $-H_W$ at a given bare quark mass $m$.
Fig. 1 shows four example plots of low lying eigenvalue densities
from the paper of \cite{EHN98}. At small $m$ the density is zero
and then jumps at a non-zero value at the critical mass $m_{cr}$
where the Wilson operator becomes singular. Then as $m$ increases
one can identify a threshold $m_{\mu}$
where the density approaches a plateau.
\begin{figure}
\vspace{4cm}
\epsfxsize=6cm
\hspace{3.3cm} \epsffile[240 400 480 450]{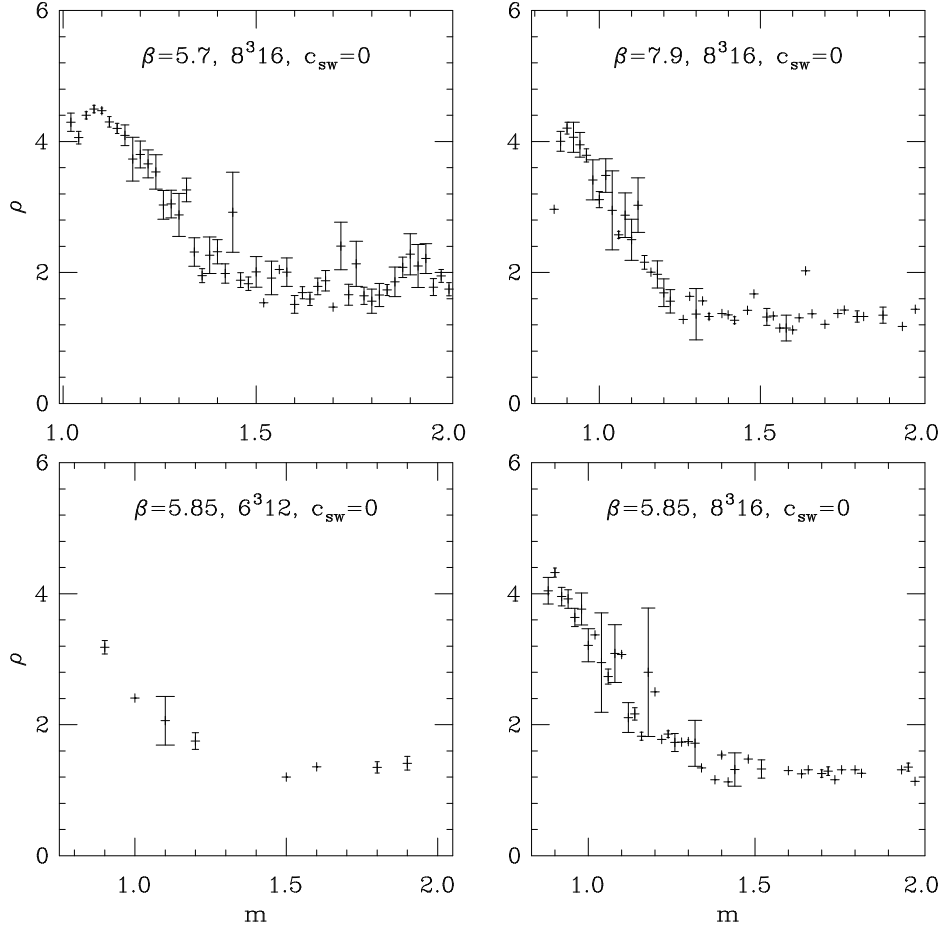}
\vspace{8cm}
\caption{Unnormalized density of zero eigenvalues $\rho(0,m)$ of $-H_W$ as
a function of the bare quark mass $m$ on $50$ quenched configurations
as computed by \cite{EHN98}.
Figure taken from the above paper using its
{\tt lanl.arXiv.org e-Print archive, hep-lat/9802016} version.
}
\end{figure}
It is this plateau where the UV effects dominate the eigenvalue
spectrum. According to this heuristics one can suppress the eigenvalues
beyond $m_{\mu}$ and the value of $\mu$ can be determined by:
\begin{equation}\label{mu_def}
\mu = m_{\mu} - m_{cr}
\end{equation}
From Fig.1 one can estimate $\mu \approx 0.3$.

\subsection{Example of UV suppression}

The effective action of a fermion theory can be written as:
\begin{equation}\label{eff_action}
S_{\text{eff}} = \text{tr} A
\end{equation}
where $A = f(H_W^2)$ is a Hermitian matrix and
$f(s)$ is a real and smooth function of $s \in \Real^{+}$.
To see the effects of UV suppression one can compute the change
of the effective action between two background $SU(3)$ gauge fields.
Since it is difficult to compute the trace directly one can use
the noisy estimators of the type $X \equiv z^T A z$
where $z$ is a ${\mathbb Z}_2$ niose vector.
It is easy to show that the random variable $X$ has
expectation value ${\mathbb E}(X) = \text{tr} A$ and variance:
\begin{equation*}
\text{Var}(X) = 2 \sum_{i \neq j} (\text{Re} a_{ij})^2
\end{equation*}
Note that bilinear forms of the type $z^T A z$
can be computed using Lanczos based methods as described by
\cite{Bai_et_al,Cahill_et_al,Borici_UVSFb}. I have used the
Lanczos algorithm as it appears in the paper of \cite{Borici_UVSFb}.
Fig. 2 shows the distribution of
$X$ and its variation $\Delta X$ between two
configurations for Wilson and UV-suppressed fermions, i.e.
$f(s) = \log \sqrt{s}, s > 0$ and $f(s) = \log \tanh \sqrt{s}/\mu, s >0$
respectively.
\begin{figure}
\vspace{4.5cm}
\epsfxsize=6cm
\hspace{3.3cm} \epsffile[240 400 480 450]{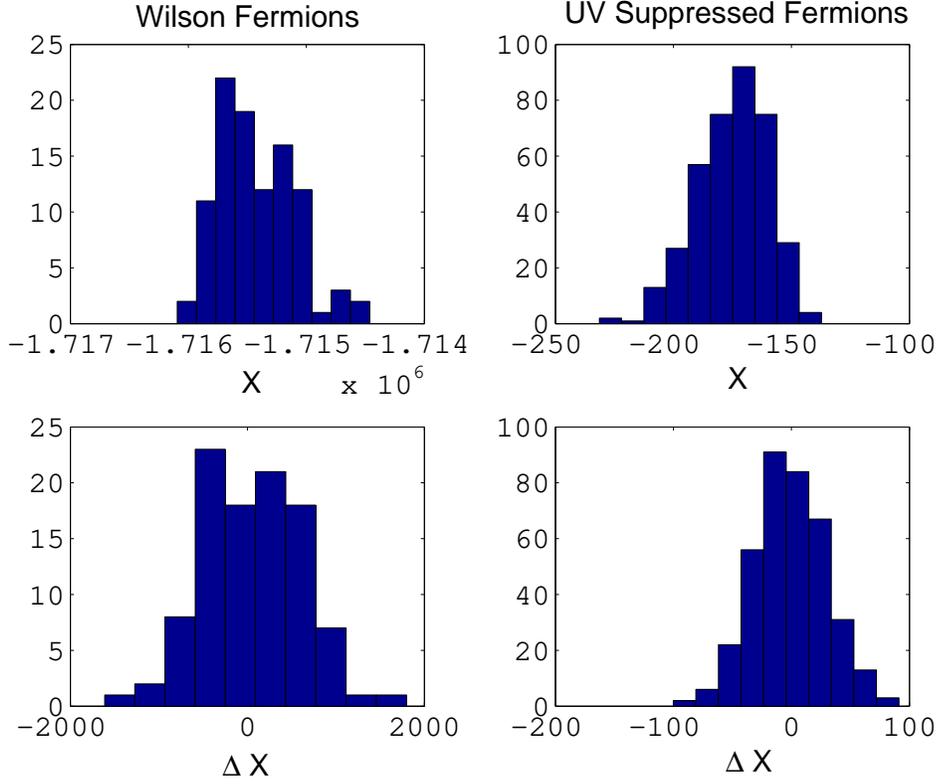}
\vspace{5.5cm}
\caption{Upper panel: Distribution of the noisy estimator $X$ of the
Wilson (left) and UV-sppressed (right) effective action. Lower panel:
The same as in the upper panel but for the variation $\Delta X$
between two configurations. Lattice parameters: $12^3\times24$,
$\beta = 5.9$, $m = -0.869$ ($\kappa = 0.1597$), $\mu = 0.2$.
}
\end{figure}
The figure shows clearly that changes $\Delta X$ in the effective
action estimator are reduced by an order of magnitude for UV-suppressed
fermions as compared to Wilson fermions.

\section{Concluding remarks}

In this paper a lattice theory with suppressed
cutoff modes of the fermion determinant is proposed.
The level of suppression can be
tuned by a new parameter $\mu$ that enters the theory.
For $\mu \rightarrow \infty$ one recovers Wilson fermions
whereas for $\mu \rightarrow 0$ one gets the quenched theory.
For general $\mu$ the theory is similar to a Wilson theory with
the effective cutoff $\sim \mu/a$.

By fixing $\mu$ at the appropriate value one can reduce
considerably the UV fluctuations in the fermion determinant
while preserving the interesting infrared modes of the theory.
Thus one can eleminate the need to separate and then devise
algorithms for the UV modes.

The fermion thoery described here can be easily extended to other
input fermions such as staggered or chiral fermions.

\vspace{2cm}

\section*{Appendix A: Proof of the result (\ref{Appendix_A})}

I follow the analogous arguments as in Appendix C of
\cite{Hernandes_et_al}.
Using the definition of the Wilson operator (\ref{Wilson_op}),
a straightforward computation yields:
\begin{equation*}
\begin{array}{l}
A(U) = D(U)_W^{\dag} D(U)_W = \\
\\
+m^2 \1 - ma \sum_{\mu = 1}^4 \nabla_{\mu} \\
\\
+ \sum_{\mu = 1}^4 (-\nabla_{\mu}^2 + \frac{a^2}{4} \Delta_{\mu}^2)
+ \sum_{\mu \neq \nu} \frac{a^2}{4} \Delta_{\mu} \Delta_{\nu}\\
\\
+ \frac{1}{2} \sum_{\mu \neq \nu} \{-\gamma_{\mu} \gamma_{\nu}
[\nabla_{\mu},\nabla_{\nu}] + a \gamma_{\mu}[\nabla_{\mu},\Delta_{\nu}]\}
\end{array}
\end{equation*}
where
\begin{equation*}
\begin{array}{l}
4a^2 [\nabla_{\mu},\nabla_{\nu}] \psi(x) =\\
\\
+[U(x)_{\mu}U(x+ae_{\mu})_{\nu} - U(x)_{\nu}U(x+ae_{\nu})_{\mu}]
\psi(x+ae_{\mu}+ae_{\nu})\\
\\
+\text{3 similar terms}
\end{array}
\end{equation*}
and
\begin{equation*}
\begin{array}{l}
2a^3 [\nabla_{\mu},\Delta_{\nu}] \psi(x) =\\
\\
+[U(x)_{\mu}U(x+ae_{\mu})_{\nu} - U(x)_{\nu}U(x+ae_{\nu})_{\mu}]
\psi(x+ae_{\mu}+ae_{\nu})\\
\\
+\text{3 similar terms}
\end{array}
\end{equation*}
Note also that:
\begin{equation*}
U(x)_{\mu}U(x+ae_{\mu})_{\nu} - U(x)_{\nu}U(x+ae_{\nu})_{\mu}
= \1 - U_{\mathbb{P}}
\end{equation*}
and
\begin{equation*}
\norm{a^2 [\nabla_{\mu},\nabla_{\nu}]}_2 = \norm{\1 - U_{\mathbb{P}}}_2, ~~~~~
\norm{a^2 [\nabla_{\mu},\Delta_{\nu}]}_2 = 2\norm{\1 - U_{\mathbb{P}}}_2
\end{equation*}
Therefore, I obtain:
\begin{equation*}
\norm{A(U)}_2 \leq \norm{D(\1)_W}_2^2
+ \frac{3}{2} \sum_{\mu \neq \nu} \norm{\1 - U_{\mathbb{P}}}_2
\end{equation*}
One can easily show that:
\begin{equation*}
\norm{\1 - U_{\mathbb{P}}}_2 \leq \norm{\1 - U_{\mathbb{P}}}_F
= \sqrt{6 [1 - P(U)]}
\end{equation*}
where $\norm{.}_F$ is the Frobenius (Euclidean) norm of a matrix.
From the definition (\ref{non_smoothness}) and
$\norm{D(\1)_W}_2 \leq m + \frac{8}{a}$,
$\norm{\delta A(U)}_2 \leq 2\norm{A(U)}_2$
(assuming that $\norm{A(U+\delta U)}_2 \leq \norm{A(U)}_2$),
I obtain the result (\ref{Appendix_A}) with:
\begin{equation*}
c_1 = 2(m + \frac{8}{a})^2, ~~~~~~c_2 = 36\sqrt{6}
\end{equation*}

\section*{Appendix B}

To prove the statement I use similar algebraic
manipulations to those used elsewhere in a different
context \cite{Borici}.
Here they appear in greater detail.

I let the lattice spacing in the fifth direction to be:
\begin{equation*}
a_5 = \frac{L_5}{n}
\end{equation*}
where $n$ is the number of lattice points in the fifth direction.
The approximate fermion measure density can be defined by:
\begin{equation*}
\omega_n(H_W) = \frac{\det(\1 - T^n)}{\det(\1 + T^n)}
\end{equation*}
with $T$ being a classical transfer matrix:
\begin{equation*}
T = \1 - a_5 H_W
\end{equation*}
It is easy to see that for small lattice spacing $a_5$, $\omega_n(H_W)
\rightarrow \omega(H_W)$.
It is only necessary to show that:
\begin{equation*}
\det(D_W^{(5)}) \sim \det(\1 - {\tilde T}^n)
\end{equation*}
where ${\tilde T} \rightarrow T$ as $a_5 \rightarrow 0$.
The right hand side can be realized for example as the determinant
of the following $n \times n$ block matrix:
\begin{equation}\label{tilde_T}
\begin{pmatrix} \1          & -{\tilde T} &        &             \\
                            & \1          & \ddots &             \\
                            &             & \ddots & -{\tilde T} \\
                -{\tilde T} &             &        & \1          \\
\end{pmatrix}
\end{equation}
where the sign of the left lower corner reverses if the boundary
conditions of $D_W^{(5)}$ change from periodic to antiperiodic.
Therefore, the ratio of the two determinants will be given by:
\begin{equation*}
\frac{\det(\1 - {\tilde T}^n)}{\det(\1 + {\tilde T}^n)}
\rightarrow \frac{\det(\1 - T^n)}{\det(\1 + T^n)} = \omega_n(H_W)
\end{equation*}

One must now calculate ${\tilde T}$ from $D_W^{(5)}$. The fermion
matrix can be written as an $n \times n$ block partition in the
fifth dimension:
\begin{equation*}
D_W^{(5)} =
\begin{pmatrix} a_5 D_W - \1 & P_+          &        & P_-          \\
                P_-          & a_5 D_W - \1 & \ddots &              \\
                             & \ddots       & \ddots & P_+          \\
                P_+          &              & P_-    & a_5 D_W - \1 \\
\end{pmatrix}
\end{equation*}
where $P_{\pm}$ are the usual spin projector operators in the
fifth direction. I multiply the above matrix from the left with
the following permutation matrix:
\begin{equation*}
\begin{pmatrix} P_+ & P_- &        &     \\
                    & P_+ & \ddots &     \\
                    &     & \ddots & P_- \\
                P_- &     &        & P_+ \\
\end{pmatrix}
\end{equation*}
and I obtain the following result:
\begin{equation*}
\gamma_5
\begin{pmatrix} a_5 P_+ H_W - \1 & a_5 P_- H_W + \1 &        &
\\
                                 & a_5 P_+ H_W - \1 & \ddots &
\\
                                 &                  & \ddots & a_5 P_-
H_W + \1 \\
                a_5 P_- H_W + \1 &                  &        & a_5 P_+
H_W - \1 \\
\end{pmatrix}
\end{equation*}
Comparing this matrix to that containing ${\tilde T}$
(\ref{tilde_T}) I arrive at the following expression for the
transfer matrix:
\begin{equation*}
{\tilde T} = \frac{\1}{\1 - a_5 P_+ H_W} (\1 + a_5 P_- H_W)
\end{equation*}
which goes to $T^{-1}$ for small lattice spacing $a_5$. Since $T$
and $T^{-1}$ are equivalent transfer matrices the proof
is concluded. \qed

\section*{Appendix C: Locality in the presence of gauge fields}

To show locality of $D$ it is sufficient to show that the
matrix-valued function $f(.)$ (\ref{f_definition})
is a local operator.

Let $p-$norm of the vector $v \in \Real^m, m \in \mathbb N$ be
defined by:
\begin{equation*}
\norm{v}_p = (\sum_{k=1}^m |v_k|^p)^{\frac{1}{p}}
\end{equation*}
with $p > 0$. The {\it induced matrix norm} of $A \in
\Real^{m\times m}$ is defined by:
\begin{equation*}
\norm{A}_p = \max_{\norm{v}_p = 1} \norm{A v}_p
\end{equation*}

If $x,y$ are two lattice points in four dimensions, I will refer
to the matrix element $(x,y)$ of the matrix function $f$ as
$f(x,y)$. Note that these elements are in fact matrices which
carry spin and color indices, which I suppress for clarity.

I will show in the following that:
\begin{equation}\label{exp_bound}
\norm{f(x,y)}_2 \leq c_1 c_2^{\frac{1}{a}\norm{x-y}_1},
~~~~~c_1 > 0, ~~~~~0 \leq c_2 < 1
\end{equation}
for any $\mu > 0$ and four dimensional lattice points $x,y$.

In order to prove (\ref{exp_bound}) I need first
to approximate the function $f$. From Appendix B one can
infer that a five dimensional formulation gives a fermion
measure density proportional to $\1 - {\tilde T}^n$. Therefore one
can write:
\begin{equation*}
D \approx \frac{1}{a} \gamma_5 \frac{\1 - {\tilde T}^n}{\1 +
{\tilde T}^n}
\end{equation*}

This suggests the following approximation to the function $f$:
\begin{equation*}
f(z^2) \approx f_n(z^2) := \frac{1}{z} \frac{(1 +
\frac{z}{2n})^{2n} - (1 - \frac{z}{2n})^{2n}}
     {(1 + \frac{z}{2n})^{2n} + (1 - \frac{z}{2n})^{2n}}
\end{equation*}

The right hand side can be expressed as a partial fraction by
computing its poles and corresponding residues:
\begin{equation}\label{f_n}
f_n(z^2) = \frac{1}{2n^2} \sum_{k=1}^n \frac{1}{(\frac{z}{2n})^2
\cos^2 \frac{\pi}{2n}(k - \frac{1}{2})
                         +\sin^2 \frac{\pi}{2n}(k - \frac{1}{2})}
\end{equation}

Now I consider the matrix valued function $f_n$ with the matrix
$\frac{aH_W}{\mu}$ substituted for the variable $z$ and show first
that it is local for any approximation order $n$.

To simplify the notation, I call $A_k$ the
matrix:
\begin{equation}\label{A_k}
A_k = (\frac{aH_W}{2n\mu})^2 \cos^2 \frac{\pi}{2n}(k -
\frac{1}{2})
                         +\sin^2 \frac{\pi}{2n}(k - \frac{1}{2})
\end{equation}
and write:
\begin{equation*}
f_n[(\frac{aH_W}{\mu})^2] = \frac{1}{2n^2} \sum_{k=1}^n
\frac{1}{A_k}
\end{equation*}

I use a geometric series to approximate the inverse of $A_k$
in order to keep the discussion simple
(however note that a Chebyshev approximation is more
accurate and is likely to decrease the exponent base $c_2$ of
(\ref{exp_bound}) in the final result \cite{Hernandes_et_al}). I
may write:
\begin{equation}\label{one_over_Ak}
\frac{1}{A_k} = \frac{1}{\norm{A_k}_2} \frac{1}{\1 - (\1 -
\frac{A_k}{\norm{A_k}_2})}
\end{equation}
and then define:
\begin{equation*}
\rho_k = \norm{\1 - \frac{A_k}{\norm{A_k}_2}}_2
\end{equation*}
From the definition of the 2-norm I obtain:
\begin{equation*}
\rho_k = 1 - \frac{1}{\norm{A_k}_2 \norm{A_k^{-1}}_2} = 1 -
\frac{1}{\kappa(A_k)}
\end{equation*}
where $\kappa(A_k)$ is the condition number of  $A_k$. It is clear
that $\rho_k, k=1,\ldots,n$ are non-negative ($\kappa(A_k) \geq
1$).

Let $\sigma_1 \leq \sigma_2$ be the extreme singular values of $a
D_W$. Then $\rho_k$ can be written as:
\begin{equation*}
\rho_k = 1 - \frac{ (\frac{\sigma_1}{2n\mu})^2 \cos^2
\frac{\pi}{2n}(k - \frac{1}{2})
                                 +\sin^2 \frac{\pi}{2n}(k -
\frac{1}{2})
     }{(\frac{\sigma_2}{2n\mu})^2 \cos^2 \frac{\pi}{2n}(k -
\frac{1}{2})
                                 +\sin^2 \frac{\pi}{2n}(k -
\frac{1}{2})
     }
\end{equation*}
or
\begin{equation*}
\rho_k = \frac{ 1 - \frac{\sigma_1^2}{\sigma_2^2} }
              { 1 + (\frac{2n\mu}{\sigma_2})^2 \tan^2 \frac{\pi}{2n}(k
- \frac{1}{2}) }
\end{equation*}
But since $\tan z > z$ for $0 < z < \frac{\pi}{2}$ (the range of
$\tan$ argument values for $k=1,\ldots,n$) I obtain:
\begin{equation*}
\rho_k < \frac{ 1 - \frac{\sigma_1^2}{\sigma_2^2} }
              { 1 + \frac{[\mu\pi(k - \frac{1}{2})]^2}{\sigma_2^2} }
\end{equation*}
Clearly, I have:
\begin{equation}\label{rho_1}
\rho_k < \rho_1 < 1, ~~~~k = 2, \ldots, n
\end{equation}

Now I may expand the right hand side of (\ref{one_over_Ak}) in
geometric series to obtain:
\begin{equation*}
\frac{1}{A_k} = \frac{1}{\norm{A_k}_2} [\1 + (\1 -
\frac{A_k}{\norm{A_k}_2}) + 
(\1 - \frac{A_k}{\norm{A_k}_2})^2 + \cdots]
\end{equation*}
Further, I let the matrix $M_k$ to be:
\begin{equation*}
M_k = (\1 - \frac{A_k}{\norm{A_k}_2})/\rho_k
\end{equation*}
such that the matrix elements of $f_n(.)$ can be given by:
\begin{equation*}
f_n(x,y) = \frac{1}{2n^2} \sum_{k=1}^n
\frac{1}{\norm{A_k}_2} \sum_{l
\geq 0} \rho_k^l M_k^l(x,y)
\end{equation*}
Since $A_k(x,y)$ vanishes for those lattice points $x,y$ such that
$\norm{x-y}_1 > 2a$, one may conclude that:
\begin{equation}\label{Ml_zero}
M_k^l(x,y) = 0, ~~~~\text{for}
~~~~\norm{x-y}_1 > 2la, ~~~~l = 1, 2, \ldots
\end{equation}
In general I have:
\begin{equation}\label{Ml_leq1}
\norm{M_k^l(x,y)}_2 \leq 1,  ~~~~l = 1, 2, \ldots
\end{equation}
Using (\ref{rho_1}), (\ref{Ml_zero}) and (\ref{Ml_leq1}), I obtain:
\begin{equation*}
\norm{f_n(x,y)}_2 < \rho_1^{\frac{1}{a}\norm{x-y}_1}
f_n[(\frac{\sigma_2}{\mu})^2]
\end{equation*}
But $z f_n(z^2) \leq 1$ for $z > 0$. Hence finally:
\begin{equation}\label{my_bound}
\norm{f_n(x,y)}_2 < \frac{\mu}{\sigma_2}
\rho_1^{\frac{1}{a}\norm{x-y}_1}
\end{equation}
Since the left hand side is uniformly bounded the result also holds
in the limit $n \rightarrow \infty$. Hence the hypothesis
(\ref{exp_bound}) is proven with the constants $c_1,c_2$ given by:
\begin{equation}\label{c1_c2}
c_1 : = \frac{\mu}{\sigma_2}, ~~~~\text{and} ~~~~c_2 : = \frac{ 1 -
\frac{\sigma_1^2}{\sigma_2^2} }
             { 1 + \frac{\mu^2 \pi^2}{4 \sigma_2^2} }
\end{equation}
\qed

{\it Remark}. If the Wilson Dirac operator is singular (at
$\sigma_1=0$, or at the ``critical hopping parameter'') the
locality of the theory is guaranteed solely by the positivity of
$\mu$.

\section*{Acknowledgements}
I would like to thank Philippe de Forcrand and Alan Irving
for discussions and useful suggestions at different stages of this work.

\vspace{2cm}


\end{document}